\documentclass[12pt]{article}

\begin{document}
\title{Molecular-Coherent-States and Molecular-Fundamental-States}
\author{Mich\`{e}le IRAC-ASTAUD \\
Laboratoire de Physique Th\'{e}orique de la mati\`{e}re
condens\'ee\\ Universit\'{e} Paris VII\\2 place Jussieu F-75251
Paris Cedex 05, FRANCE\\
e-mail : mici@ccr.jussieu.fr}
\date{}
\maketitle

to Mosh\'{e} Flato and Andr\'{e} Heslot.

\begin{abstract}
New families of Molecular-Coherent-States are constructed by the
Perelomov group-method. Each family is generated by a
Molecular-Fundamental-State that depends on an arbitrary sequence
of complex numbers $c_j$. Two of these families were already
obtained by D.Janssen and by J. A.
 Morales, E. Deumens and Y. \"{O}hrn.
The properties of these families are investigated and we show that
most of them are independent on the $c_j$.
\end{abstract}




\maketitle

\section{Introduction}

Since their introduction by Schr\"odinger \cite{schrodinger}, the
Coherent States of the Harmonic Oscillator (C.S.H.O.) were
extensively studied and used in many branches of physics
\cite{Klauder}. These states satisfy numerous properties, let us
recall some of them~:

$\bullet $ 1) The C.S.H.O. constitute an (overcomplete) basis of
non orthogonal vectors of the Hilbert Space of the states of the
harmonic oscillator ${\cal H}$.

$\bullet $ 2) On this basis, the vectors of ${\cal H}$ are realized
as entire analytical functions of a complex variable.

$\bullet $ 3) The C.S.H.O. are eigenvectors of the annihilation
operator.

$\bullet $ 4) They minimize the uncertainty relations.

$\bullet $ 5) The mean values of the position and of the momentum
on the C.S.H.O. evolve in time like the corresponding classical
quantities.

$\bullet $ 6) The C.S.H.O. have the temporal stability.

$\bullet $ 7) They are generated by
 the Heisenberg-Weyl group.

 Their
generalizations to others systems are constructed in order to
verify some of these properties, 1 and 2 being always required.

A fruitful generalization originating from the property 7 was given
by Perelomov who defined
  Coherent States related to other
Lie Groups \cite{perelomov}. Applying the group-method to $SU(2)$
\cite{perelomov1}, he constructed the Spin-Coherent-States studied
by Radcliffe \cite{radcliffe}\cite{arecchi}.

In \cite{janssen} and \cite{mdo}, Coherent-States ( C.S.) were
found for the quantum mechanical top and for the description of
molecular-rotations. These states fulfill the two requirements 1
and 2 and are proved to satisfy the properties 4 and 5. We claim
that the proof, based on some relations satisfied by these C.S., is
not valid because the property 6 is not fulfilled, i.e. a rotor in
a C.S. introduced in \cite{janssen} and \cite{mdo} does not remain
in a C.S. when time evolves. We come back in detail on this point
in the following.

The main interest of the C.S. introduced in \cite{janssen} and
\cite{mdo} and of the Spin-Coherent-States is to constitute a
suitable basis in various applications~: asymmetric top
\cite{pavlichenkov}, forced rotation model \cite{janssen}, time
dependent electron nuclear dynamics \cite{mdo}, partition function
in a magnetic field \cite{radcliffe}, spin relaxation process
\cite{takahashi}$\cdots$ The C.S. are not unique and the purpose of
this paper is to construct
 new families of C.S. generalizing the
states introduced in \cite{janssen} and \cite{mdo}, to study and
compare their properties.

 To begin, in Section
(\ref{Angular}), we recall some well-known properties of the
quantum rigid body in order to fix the notations. In section
(\ref{coherent}), we define Molecular-Coherent-States (denoted in
the following M.C.S.) as the result of the action of
group-operators on a Molecular-Fundamental State (denoted in the
following M.F.S.). The Lie Group acting is $SU(2)\otimes SU(2)$ and
the M.F.S. is a generalization of the fundamental vector used in
\cite{perelomov}. A M.F.S. is characterized by a sequence of
complex numbers $c_j$, that must verify two conditions in order
that the M.C.S. satisfy the requirements 1 and 2. A large
arbitrariness remains in the choice of the M.F.S., but once this
choice is done, the set of M.C.S. is uniquely defined. The C.S.
defined in \cite{janssen} and \cite{mdo} are recovered for two
specific sequences of $c_j$.

In section (\ref{prop-BS}), we give the characteristic properties
of the M.F.S., the main result is that the choice of the $c_j$ does
not play a prominent part in this study.

In section (\ref{prop-CS}), all the results of the previous section
are transformed by the action of the group to set up the list of
the characteristic properties of the M.C.S.. The $Z$-representation
is tackled and the representation of the angular momentum as
differential operators is given.

The conclusions are contained in the last section and the
appendices give some complements : the realization of the angular
momentum on the functions of the Euler angles and the
representation of the bi-tensors on the canonical basis.

\section{Quantum Rigid Body}\label{Angular}

\subsection{Laboratory and Molecular-Components of the Angular Momentum }

 For any quantum system, the
components $J^L_{(0,1,2)}$ of its angular momentum $\vec{J}$, on a
set of three mutually orthogonal laboratory-fixed axes are the
generators of the rotation group and verify the $su(2)$-algebra
commutation relations~:

\begin{equation}\label{com'}
\left[J^L_{+} ,J^L_{-}\right] = 2  J^L_{0}, \quad
\left[J^L_{0}, J^L_{\pm}\right] = \pm  J^L_{\pm}
\end{equation}
where the spherical coordinates are defined by $J_{\pm}
\equiv J_1
\pm i J_2$.

The top-hamiltonian reads

\begin{equation}\label{rotor}
H= \sum_{i=0}^{2} A_i (J^M_i)^2
\end{equation}
where $J^M_{(0,1,2)}$ are the components of the angular momentum
$\vec{J}$ on a set of three mutually orthogonal axes moving with
the system under a rotation. The rotational constants $A_{i}$,
inverse of the moments of inertia, characterize the symmetry of the
molecule. Writing the molecular-components $J^M$ as the scalar
product
 of $\vec{J}$ with a vector and  using the characteristic commutation relations
 of $J^L $ with the laboratory-components of a vector, we easily prove
that the $J^M$ satisfy the following commutation relations

\begin{equation}\label{com}
\left[ J^M_+,J^M_-\right] = -2 J^M_0, \quad
\left[J^M_0, J^M_\pm \right] = \mp  J^M_\pm
\end{equation}
and that they commute with all the laboratory-components, symbolically~:

\begin{equation}\label{JJ}
\left[J^L,J^M\right]=0
\end{equation}
The Lie algebra ${\cal A}$, generated by the $J^L$ and the $J^M$ is
$su(2)\otimes su(2)$ with the constraint $J^2=
\sum_{i=0}^{2}(J^L_i)^2= \sum_{i=0}^{2}(J^M_i)^2$.

Up to now, the rotations considered are the rotations of the body
and of the molecular-frame that keep fixed the laboratory-frame;
they correspond to
 unitary transformations $R_L$ of the
states and observables of the quantum system

\begin{equation}\label{rot-L}
 R_L (\alpha_L,\beta_L, \gamma_L) = \exp(-i\alpha_L
J^L_{0})\exp(-i\beta_L J^L_{2})\exp(-i\gamma_L J^L_{0})
\end{equation}
Similarly, we can consider rotations of the body and of the
laboratory-frame that keep fixed the molecular-frame; the unitary
operator $R_M$ associated to these rotations are given by

\begin{equation}\label{rot-M}
 R_M (\alpha_M,\beta_M, \gamma_M) = \exp(-i\alpha_M
J^M_{0})\exp(i\beta_M J^M_{2})\exp(-i\gamma_M J^M_{0})
\end{equation}

Due to (\ref{JJ}), the laboratory-rotations, $R_L$,
and the molecular-rotations, $R_M$, commute.

\subsection{ Representation of $J^L$ and $J^M$}

The eigenvectors of the three operators $J^2, J^L_0, J^M_0$
constitute the basis of the space of the canonical representation.
We have~:

\begin{equation}\label{Jkm}
  \begin{array}{llll}
    J^2 \mid j, k, m >&= j(j+1) \mid j, k, m >,& j =& 0,\frac{1}{2},1, \cdots\\
    J^L_{z}\mid j, k, m >& = m \mid j, k, m >,& m  =& -j,-j+1, \cdots,j\\
J^M_{z}\mid j, k, m >& = k \mid j, k, m >,& k =& -j,-j+1, \cdots,j
  \end{array}
\end{equation}
When $j$ is fixed, the states $\mid j, k, m >$
   span the $(2j+1)^2$dimensional Hilbert space $h_j$.

The action of the operators on the canonical basis is given
by~:

\begin{equation}\label{reprJJ'}
\begin{array}{ll}
  J^L_{\pm}\mid j, k, m >
  &=\sqrt{(j\mp m)(j\pm m +1)} \mid j, k, m \pm 1>\\
  &\\
    J^M_{\pm}\mid j, k, m >
    &=\sqrt{(j\pm k)(j\mp k +1)} \mid j, k\mp 1, m >\\
\end{array}
\end{equation}

Let $\alpha,
\beta,\gamma$ be the Euler angles relating
 the laboratory-frame and the molecular-frame and satisfying, by convention,

\begin{equation}\label{Euler}
    0 \leq \alpha < 2 \pi, \quad
    0 \leq \beta \leq  \pi, \quad
    -\pi \leq \gamma <  \pi.
\end{equation}
 It is well known that the canonical representation can be realized on  the space of
functions, ${\cal C}(\alpha,
\beta,\gamma)$, and that $j$ must be an integer number in order
that
 the wave functions of a rigid molecule  be
single valued (see appendix (\ref{ap-Euler})).

 In
the following, we don't restrict to this case and we construct the
M.C.S. spanning either ${\cal{H}}_\frac{1}{2}
\equiv
\oplus_ {(j = 0,\frac{1}{2},1, \cdots)} h_j$ or ${\cal{H}}_1
\equiv
\oplus_ {(j = 0,1, \cdots)} h_j$.

\subsection{Bi-tensors}

 The components of a bi-tensor operator  commute between themselves and
 transform under the laboratory or
  the molecular-rotations according to the formulas \cite{judd}

\begin{equation}\label{tens-R}
R_L T^{j, j'}_{q,q'}R_L^{-1} = \sum _{k'=-j'}^{j'} T^{j, j'}_{q,
k'} R^{j'}_{k' q'}, \quad R_M T^{j, j'}_{q, q'}R_M^{-1} = \sum
_{k=-j}^{j} T^{j, j'}_{k, q'} R^{j}_{k q}
\end{equation}

where

\begin{equation}\label{Djmm'1}
  R^j_{mm'}(\alpha,\beta,\gamma ) =\exp(-i\alpha m)\exp(-i\gamma m') d^j_{mm'}(\beta)
\end{equation}

and

\begin{equation}\label{Djmm'2}
  \begin{array}{ll}
d^j_{m'm}(\beta)\equiv&
\sqrt{(j-m')!(j+m')!(j-m)!(j+m)!}
\left(\tan\left(\frac{\beta}{2}\right)\right)^{m-m'}
\left(\cos\left(\frac{\beta}{2}\right)\right)^{2j}\\
&\times \sum
\frac{(-1)^n}{n!(j-m-n)!(j+m'-n)!(m-m'+n)!}
\left(\tan\left(\frac{\beta}{2}\right)\right)^{2n}
\end{array}
\end{equation}

The resulting commutation relations read:

\begin{equation}\label{J'T}
\begin{array}{llll}
\left[ J^L_\pm , T^{j, j'}_{q, q'}\right ]
&=\sqrt{(j'\mp q')(j'\pm q' +1)}T^{j, j'}_{q,q'\pm 1}, \quad
&\left[ J^L_0 , T^{j, j'}_{q, q'} \right ] &= q' T^{j, j'}_{q,
q'}\\
&&&\\
\left[ J^M_\mp , T^{j, j'}_{q, q'} \right]
&=\sqrt{(j\mp q)(j \pm q +1)}T^{j, j'}_{q\pm 1, q'}, \quad &\left[
J^M_0 , T^{j, j'}_{q, q'} \right] &= q T^{j, j'}_{q, q'}
\end{array}
\end{equation}

The hermitean adjoint is defined by

\begin{equation}\label{her}
(T^\dagger)^{j j'}_{q q'}= (-1)^{q-q'} T^{j j'}_{-q -q'}
\end{equation}

We easily verify that $(-\frac{1}{\sqrt{2}}J^L_+, J^L_0,
\frac{1}{\sqrt{2}}J^L_-) $ is a bi-tensor $J^{0,1}$ and that
 $(-\frac{1}{\sqrt{2}}J^M_-, J^M_0,
\frac{1}{\sqrt{2}}J^M_+)$ is a bi-tensor $J^{1,0}$.

In the following, we call bi-spinor $S$ the bi-tensor
$T^{\frac{1}{2},\frac{1}{2}}$ and bi-vector $V$ the bi-tensor
$T^{1,1}$. The components of $S$ and $V$ are represented on the
canonical
basis  in the Appendix \ref{S-V}.

\section{Coherent States}\label{coherent}

\subsection{Definitions}

{\bf Definition} Let $c_0 \ne 0,c_{\frac{1}{2}}, c_1
\cdots ,$ be an arbitrary sequence of complex numbers,
 a Molecular-Fundamental-State (M.F.S.), is a state of ${\cal{H}}_\frac{1}{2}$ of the
 form~:

\begin{equation}\label{initial}
\mid z > \equiv \sum_j c_j
z^j \mid j,-j,-j>, \quad z \in C,
 \quad \sum_j \equiv \sum
_{j=0,\frac{1}{2},1 \cdots}
\end{equation}
The M.F.S. is the analogous of the fundamental vector generating
the Spin C.S. in \cite{perelomov} and constitutes the main
ingredient of the group-construction of the M.C.S. proposed in this
paper. In order that the M.F.S. belongs to the Hilbert space
${\cal{H}}_\frac{1}{2}$, the coefficients $c_j$ and the complex
variable $z$ must satisfy the following condition

\begin{equation}\label{cond1-cj}
<z\mid z> \equiv N(\mid z \mid^2) = \sum_j \mid c_j\mid^2 \mid z
\mid^{2j}<
\infty
\end{equation}
Let us remark that $\mid z>$ belongs to ${\cal{H}}_1$, if $c_j = 0$
when $j$ take half-integer values.

Applying the group-method \cite{perelomov} to the group $SU(2)
\otimes SU(2)$, we define

{\bf Definition} The Molecular-Coherent-States are the states
resulting from the action of the laboratory-rotations $R_L$ and of
the molecular-rotations $R_M$ upon the M.F.S.(\ref{initial}) and
spanning ${\cal{H}}_\frac{1}{2}$ (or ${\cal{H}}_1$ when $\mid z>
\in {\cal{H}}_1$).

Let us denote

\begin{equation}\label{DL}
D_L(\zeta_L) = e^{\zeta_L J^L_+}e^{\eta_L J^L_0}
e^{-\overline{\zeta_L}J^L_-}, \quad
    \eta_L = \ln(1+\mid \zeta_L \mid^2)
\end{equation}
Writing $R_L (\alpha_L,\beta_L, \gamma_L)$ as the product of $D_L(-
\tan \frac{\beta_L}{2}e^{-i\alpha_L})$ by
$e^{-i(\alpha_L+\gamma_L)J^L_0}$, we notice that the last term of
this product transforms the M.F.S. $\mid z>$ into $\mid
ze^{i(\alpha_L+\gamma_L)}>$, and therefore that two
laboratory-rotations only differing by this last term give the same
M.C.S.. The same holds for the molecular-rotations.

Therefore, analogously to $SU(2)$, a M.C.S. system is constructed
by applying on the M.F.S.(\ref{initial}) the operators $D_L$
defined in (\ref{DL}) and $D_M$ defined by~:

\begin{equation}\label{DM}
D_M(\zeta_M) = e^{\zeta_M J^M_-}e^{\eta_M J^M_0}
e^{-\overline{\zeta_M}J^M_+}, \quad
    \eta_M = \ln(1+\mid \zeta_M \mid^2)
\end{equation}
A M.C.S. then is of the form

\begin{equation}\label{eta}
\mid Z >
 =  D_L(\zeta_L) D_M(\zeta_M)\mid z >
\end{equation}
Obviously, the norms of $\mid Z >$ and $\mid z>$ are both equal to
$\sqrt{N(\mid z\mid^2)}$ and the M.C.S. exist if the sequence $c_j$
and the complex parameter $z$ verify (\ref{cond1-cj}).

The explicit calculation of (\ref{eta}) gives the decomposition of
the M.C.S. on the canonical basis

\begin{equation}\label{eta1}
 \mid Z > =
 \sum_{jkm} \sigma^j_k \sigma^j_m
\zeta_L^{j+m} \zeta_M^{j+k}
z^j c_j (1+\mid \zeta_L \mid^2)^{-j}(1+\mid \zeta_M\mid^2)^{-j}
\mid j k m>
\end{equation}
with the notations

\begin{equation}\label{param}
\sum_{jkm} \equiv
\sum_{j=0}^\infty
\sum
_{m=-j}^j
\sum
_{k=-j}^j \quad \mbox{and} \quad \sigma^j_k \equiv \sqrt{\frac{(2j)!}{(j-k)!(j+k)!}}.
\end{equation}

{\bf Definition} A c-set is a set of M.C.S. defined by
(\ref{initial}) and (\ref{eta}) and corresponding to a given
sequence $c_j$.

 The parameter
 $Z= (z, \zeta_L, \zeta_M,)$  is  such as
$\zeta_L$ and $\zeta_M$ belong to the whole complex plane $C$, $z$
is eventually restricted by (\ref{cond1-cj}).

The scalar product of two M.C.S. of a c-set is given by

\begin{equation}\label{pro-sca}
< Z' \mid Z > = N\left( \frac{(1+\overline{\zeta'}_L\zeta_L)^2
     (1+\overline{\zeta'}_M\zeta_M)^2\overline{z'}z}
     {(1+\mid \zeta_L \mid^2)(1+\mid \zeta_M\mid^2)
 (1+\mid \zeta'_L \mid^2)(1+\mid \zeta'_M\mid^2)}\right)
\end{equation}

We illustrate each step of the present study with eight examples.

\vspace{0.5cm}
$\clubsuit$ In the following examples,
 all the  representations occur in the
decomposition of the M.C.S. over the canonical basis.

\begin{tabular}{|c|c|c|} \hline
     &$c_j$, $j$ integer or half-integer & $N(\mid z \mid ^2)$ \\
    \hline
\hline
   1 &$\frac{1}{\sqrt{(2j)!}}$ & $e^{\mid z \mid },\quad \forall \mid z \mid$  \\ \hline
    2&$\sqrt{\frac{2j+1}{2}}\frac{1}{\sqrt{(2j)!}}$ & $\frac{1}{2}(1+\mid z \mid )
   e^{\mid z \mid },\quad \forall \mid z \mid$  \\ \hline
  3& $(2j+1)\sqrt{j+1}$&$\frac{3 \mid z \mid  + 2}{2(1-\mid z \mid
   )^4}, \quad \mid z \mid < 1$\\ \hline
  4& $(2j+1)^{\frac{3}{2}}$&$\frac{\mid z \mid^2+4\mid z \mid+1}{(1-\mid z \mid)^4},\quad \mid z \mid <
   1$\\
   \hline
\end{tabular}
\vspace{0.5cm}

Let us stress that :

The M.C.S. 1 were previously studied by D.Janssen \cite{janssen}
who arbitrarily assumed the values of the $c_j$. In \cite{bhaumik},
\cite{atkins} \cite{fonda} and \cite{takahashi}, these specific
values of the $c_j$ are obtained for the linear rotor ($k=0$) by
using the Schwinger'method for the construction of the angular
momentum algebra
.

The M.C.S.3 and the M.C.S.4 only exist if $\mid z
\mid < 1$.
\vspace{0.5cm}

$\clubsuit$ In the following examples, the coefficients are such
that
$c_j = 0$
 when $j$ is an half-integer number, the M.C.S. only depend on odd-dimensional
 representations
and belong to
 ${\cal{H}}_1$.

\begin{tabular}{|c|c|c|} \hline
     &$c_j$, $j$ integer & $N(\mid z \mid ^2)$ \\
    \hline
    \hline
  5& $\frac{1}{\sqrt{j!}}$ & $e^{\mid z \mid ^2}, \quad \forall \mid z \mid$ \\
  \hline
   6& $\frac{2j+1}{\sqrt{j!}}$ & $(4 \mid z \mid ^4 +8 \mid z \mid ^2 +1)e^{\mid z \mid ^2},\quad \forall \mid z \mid$ \\
   \hline
   7&$(2j+1)\sqrt{j+1}$&$\frac{9 \mid z \mid ^4 + 14 \mid z \mid ^2 +1}{(1-\mid z \mid^2)^4},
   \quad \mid z \mid < 1$\\
   \hline
  8& $(2j+1)^{\frac{3}{2}}$&$\frac{(1+\mid z \mid^2)
   (\mid z \mid^4+22\mid z \mid^2+1)}{(1-\mid z \mid^2)^4},\quad \mid z \mid <
   1$\\
  \hline
\end{tabular}
\vspace{0.5cm}

The M.C.S.5 were previously introduced by Jorge A.
 Morales, Erik Deumens and Yngve \"{O}hrn who analyze the results occuring
  when the half-integer values of $j$ are discarded from the study of Janssen
  \cite{mdo}.

 The M.C.S.7 and the M.C.S.8 only exist if $\mid z
\mid < 1$.

\subsection{Resolution of unity}

We impose that the set of M.C.S. span ${\cal{H}}_\frac{1}{2}$
(resp. ${\cal{H}}_1$) and verify a resolution of unity

\begin{equation}\label{rel-fer}
\frac{1}{\pi^3}\int\int \int dz d\overline{z} \frac{d\zeta_L
d\overline{\zeta_L}}{(1+\mid \zeta_L \mid^2)^2}\frac{d\zeta_M
d\overline{\zeta_M}}{(1+\mid \zeta_M \mid^2)^2} f(\mid z \mid ^2)
 \mid Z >< Z\mid = 1
\end{equation}
where the operator $1$ is the unity in ${\cal{H}}_\frac{1}{2}$
(resp. in ${\cal{H}}_1$). The measures in the $\zeta_L$ and
$\zeta_M$-integrations are the $SU(2)$- invariant measures and the
measure $f(\mid z
\mid ^2)$  must be determined. The calculation of the expression
(\ref{rel-fer}) between two states $\mid j k m>$ and $\mid j' k'
m'>$ shows that the weight-function $f(x)$ is the Mellin-inverse of
the function $\hat{f}$ such as~:

\begin{equation}\label{f}
  \int dx f(x) x^j = \frac{(2j+1)^2}{\mid c_j \mid^2}\equiv
  \hat{f}(j),
  \quad j\in 2N \quad (resp. j\in N)
\end{equation}
 The existence of $f(x)$ and therefore of the
 resolution of the identity only depends on the choice of the coefficients
$c_j$.

{\bf Result :} The resolution of the identity (\ref{rel-fer})
exists if the sequence $c_j$ is such that the function $\hat{f}$
defined by (\ref{f}) is the Mellin transform of a function $f$.

A large arbitrariness remains in the choice of the $c_j$.
Reciprocally, any function leading to finite momenta gives a
sequence $c_j$ using (\ref{f})and then a family of M.C.S. provided
that the set of $z$ verifying (\ref{cond1-cj}) is not reduced to
$0$.

Let us remark that we cannot restrict the complex variable $z$ to
be on a circle because Formula (\ref{f}) then implies that $\mid
z\mid^{2j}
\mid c_j \mid^2 = (2j+1)^2 $ and then that the norm of the M.C.S. is infinite.

In the following table, we give the measure $f(x)$ corresponding to
the eight examples illustrating the construction.

\vspace{0.5cm}

\begin{tabular}{|c|c|} \hline
    & $f(\mid z \mid ^2)$ \\
    \hline
    \hline
   1 & $\frac{1}{2}(\mid z \mid-1)e^{-\mid z \mid }$  \\ \hline
   2 & $e^{-\mid z \mid }$  \\ \hline
   3 & $\theta(1-\mid z \mid)$\\
  \hline
  4 &$\frac{\theta(1-\mid z \mid)}{2\mid z
  \mid}$\\
  \hline
  \hline
  5 & $(4 \mid z \mid ^4 -8 \mid z \mid ^2 +1)
  e^{-\mid z \mid ^2}$ \\
  \hline
   6 & $ e^{-\mid z \mid ^2}$ \\
  \hline
  7 & $\theta(1-\mid z \mid)$\\
   \hline
  8 &$\frac{\theta(1-\mid z \mid)}{2\mid z
  \mid}$\\
  \hline
\end{tabular}
\vspace{0.5cm}

\noindent $\theta(x)$ is the Heaviside-function equal to $1$ when $ x >0 $ and to $0$ when $ x <0 $.
 Let us remark that the measures are strictly positive except  $f_1(x)$
 and $f_5(x)$,  previously obtained by \cite{janssen} and \cite{mdo}.

 Formula (\ref{rel-fer}) implies the
  existence of a reproducing kernel $<Z\mid Z'>$.

To conclude, due to the existence of a resolution of unity, we are
able to decompose any state $\mid \psi >$ on
  the overcomplete basis of the M.C.S.. This gives the
  $Z$-representation of ${\cal A}$.

\subsection{Z-Representation}\label{Z-Representation}

In the $Z$-Representation, an arbitrary state
 of ${\cal{H}}_\frac{1}{2}$
of the form $\mid \psi > = \sum_{jkm} c_{jkm} \mid jk m >$,
corresponds to a continuous function $\psi (Z)
\equiv <\overline{Z}\mid
\psi> $ of the three
 complex variables $\zeta,\zeta_L $
and $\zeta_M $ where $\zeta$ is defined by

\begin{equation}\label{zeta}
\zeta = \frac{\zeta_L\zeta_M z}{(1+\mid \zeta_L \mid^2)(1+\mid \zeta_M \mid^2)}
\end{equation}
Using the expression (\ref{eta}), we obtain

\begin{equation}\label{f(Z)}
  \begin{array}{ll}
\psi (Z) &= <\overline{Z}\mid \psi> \\
&= \sum_{jkm}c_{jkm}\sigma^j_m \sigma^j_k
\zeta_L^{m}\zeta_M^{k}\overline{c_j}
\zeta^j
\end{array}
\end{equation}
The function associated to the M.F.S. $\mid z_0>$ is
$N\left(\frac{z_0z}{\zeta_L\zeta_M}\right)$, where $N$ is the
norm-function introduced in (\ref{cond1-cj}). In this
representation, the components of $\vec{J}$, obtained by
calculating $<\overline{Z}\mid J \mid j,k,m
>$, don't depend on the sequence $c_j$ and take the very simple
form~:

\begin{equation}\label{oper-labo}
\left\{
  \begin{array}{ll}
    J^L_- &=
    \frac{1}{\zeta_L}(\zeta \partial_\zeta + \zeta_L\partial_{\zeta_L} ) \\
    J^L_+  &= \zeta_L(\zeta \partial_\zeta - \zeta_L \partial_{\zeta_L}) \\
    J^L_0  &= \zeta_L \partial_{\zeta_L}
  \end{array}
  \right.
\end{equation}
and

\begin{equation}\label{oper-mole}
\left\{
  \begin{array}{ll}
    J^M_+ &= \frac{1}{\zeta_M}(\zeta \partial_\zeta + \zeta_M\partial_{\zeta_M} ) \\
    J^M_-  &= \zeta_M(\zeta \partial_\zeta - \zeta_M \partial_{\zeta_M}) \\
    J^M_0  &= \zeta_M \partial_{\zeta_M}
  \end{array}
  \right.
\end{equation}
These expressions can be used to calculate the matrix elements
$<\overline{Z}\mid T \mid Z'>$ when $T$ is a polynomial of the
components of $\vec{J}$ they are obtained by differentiating the
norm-function. Let us remark that due to the specific ranges of the
parameters $j,k,m$ given in (\ref{param}), the space of the
functions occurring in (\ref{f(Z)}) is a subspace of ${\cal
C}(\zeta,\zeta_L,\zeta_M)$ and that the operators $J_+$ and $J_-$
are not adjoint in the whole space but only in the subspace.

The expression of $<\overline{Z}\mid S\mid j,k,m >$ involves
coefficients such as $\sqrt{j}, \cdots$ that correspond to
undefined operators $\sqrt{\zeta
\partial_\zeta}$ and then the bi-spinor is not a differential operator.
In the $Z$-representation, all the operators $B$ can be determined
by the diagonal elements $<\overline{Z}\mid B\mid \overline{Z}>$.
We give the expressions of these quantities when $B$ is a component
of $\vec{J}$, $S$ and $V$ in Section (\ref{prop-CS}).

To end let us give an application of the Z-representation, the
study of the asymmetric-top. Replacing (\ref{oper-mole}) in the
Hamiltonian (\ref{rotor}), we find that the stationary wave
functions satisfy a differential equation in the complex variable
$\zeta_M$, this equation was previously obtained and studied by
Pavlichenkov \cite{pavlichenkov}.

\section{Properties of the Molecular-Fundamental-States}\label{prop-BS}

The M.C.S. being obtained by the action of rotations upon the
M.F.S., the study of their properties is simpler if deduced from
the properties of the M.F.S.. Let us stress that in the following,
the sequence $c_j$ is not specified, moreover we shall see that,
except for explicit calculations,
 the specific choice of this sequence
 does not
play an important part.

\subsection{Action of the angular momentum on the M.F.S.}

From now on, we discard the label $L$ or $M$ of the components when
 the formulas  hold in both cases. Let us define the operator
 $\Lambda $  by its action on the
canonical basis

\begin{equation}
\Lambda
\mid j,k,m>= j
\mid j,k,m>
\end{equation}
  $\Lambda$ commutes
 with all the generators of ${\cal A}$.We easily prove that

 \begin{equation}\label{prop-ini1}
J_0\mid z> = -\Lambda \mid z>, \quad J^L_- \mid z> = 0, \quad
     J^M_+ \mid z>  = 0
\end{equation}
and that

\begin{equation}\label{prop-ini2}
  \begin{array}{lll}
    ( J^2 + J_0(1-J_0))\mid z> &=(J^L_-J^L_++2 J_0 )\mid z>&\\
    &=(J^M_+J^M_- +2 J_0 )\mid z>&=0
  \end{array}
\end{equation}

Let us remark that Relations (\ref{prop-ini1}) and
(\ref{prop-ini2}) do not depend on the sequence $c_j$.

\subsection{ Mean values of the angular momentum}

  We  calculate

 \begin{equation}\label{J-ini}
<z' \mid J_\pm \mid z> = <z' \mid J_{1} \mid z> = <z' \mid J_{2}
\mid z> = 0
\end{equation}
and

\begin{equation}\label{Jz-ini}
<z' \mid J_0 \mid z>
      = \sum (-j)\mid c_j \mid^2(z\overline{z'})^j= -(z\overline{z'})
      N'(z\overline{z'})
\end{equation}
The mean values of the square of the components are given by~:

\begin{equation}\label{Jx2-ini}
    <z' \mid 2 (J_1)^2 \mid z>
    = <z' \mid 2 (J_2)^2\mid z>
      =  -  <z' \mid J_0 \mid z>
\end{equation}
and

\begin{equation}\label{Jz2-ini}
<z' \mid (J_0)^2 \mid z>
= (z\overline{z'})(z\overline{z'}N''(z\overline{z'})+N'(z\overline{z'}))
\end{equation}
It results that

\begin{equation}\label{J2ini}
    <z' \mid J^2 \mid z>
=z\overline{z'}\left( z\overline{z'}N''(z\overline{z'}) +2
N'(z\overline{z'}) \right)
\end{equation}

Let $<T>_z$ be the expectation value of the operator $T$ in the
M.F.S. $\mid z>$,

\begin{equation}\label{<>}
< T >_z \equiv \frac{<z \mid T \mid z>}{<z \mid z>}
\end{equation}
 from (\ref{J-ini}) and (\ref{Jz-ini}), it results that

\begin{equation}\label{<J>}
  \begin{array}{ll}
   <J_\pm  >_z &= 0  \\
   &\\
    <J_0 >_z &= -  \mid z \mid^2 \frac{N'(\mid z \mid^2)}{N(\mid z \mid^2)}\\
    &\\
     <J^2 >_z &=   \mid z \mid^2 \left( \mid z \mid^2
     \frac{N''(\mid z \mid^2)}{N(\mid z \mid^2)}+
     2 \frac{N'(\mid z \mid^2)}{N(\mid z \mid^2)}\right)
  \end{array}
\end{equation}

{\bf Result : } The vector of components $<J^L_{0,1,2}>_z$ (resp.
$<J^M_{0,1,2}>_z$)
  lies on the $x_0$-axis
 of the laboratory-frame (resp.
 of the molecular-frame), this property is independent of the
 choice of the sequence $c_j$.

We calculate the mean values given in (\ref{<J>}) for our eight
examples. The cases 1 and 5 were previously obtained in
\cite{janssen} and \cite {mdo}.

\vspace{0.5cm}

\begin{tabular}{|c|c|c|} \hline
     & $<J_0>_z$&$<J^2>_z$ \\
\hline
  \hline
&&\\
  1 &
   $-\frac{1}{2}\mid z \mid$&
   $\frac{1}{4}\mid z \mid \left( 3+ \mid z \mid\right)$ \\ \hline
   &&\\
   2&
    $-\frac{1}{2}\mid z \mid\frac{\mid z \mid +2}{\mid z \mid +1}$
    &
     $\frac{1}{4}\mid z \mid
     \frac{\mid z \mid^2 +6 \mid z \mid +6}{\mid z \mid+1}$ \\ \hline
     &&\\
   3& $-\frac{1}{2}\mid z \mid
   \frac{9\mid z \mid +11}{\left( 1-\mid z \mid \right)
   \left( 3\mid z \mid +2\right)}$&$\frac{1}{4}\mid z \mid
     \frac{9 \mid z \mid^2 +58 \mid z \mid +33}{\left( 1-\mid z \mid
     \right)^2
   \left( 3\mid z \mid +2\right)}$\\
  \hline
   &&\\
  4&
  $-\mid z \mid\frac{\mid z \mid^2+7\mid z \mid+4}{(1-\mid z \mid)(\mid z \mid^2+4\mid z
  \mid+1)}$&$\mid z \mid\frac{6(\mid z \mid^2+3\mid z \mid+1)}{(1-\mid z \mid)^2(\mid z \mid^2+4\mid z
  \mid+1)}$\\
  \hline
    \hline
    &&\\
  5 & $-\mid z \mid ^2$&
  $\mid z \mid ^2(\mid z \mid ^2 +2)$\\
  \hline
  &&\\
   6 & $ - \mid z \mid ^2 \frac{4\mid z \mid ^4 +16 \mid z \mid ^2 +9}
   {4\mid z \mid ^4 +8 \mid z \mid ^2 +1}$&
    $ \mid z \mid ^2 (\mid z \mid ^2 +2)\frac{ 4\mid z \mid ^4 +
     +24 \mid z \mid +9}
   {4\mid z \mid ^4 +8 \mid z \mid ^2 +1}$\\
  \hline
&&\\
  7 & $ -\mid z \mid ^2\frac{6  \left(3\mid z \mid ^4 +
  10 \mid z \mid ^2 +3 \right)}
   {\left(1- \mid z \mid ^2\right)
   \left(9\mid z \mid ^4 +14 \mid z \mid ^2 +1\right)}$&
   $ \mid z \mid ^2 \frac{6  \left(3\mid z \mid ^6 +32 \mid z \mid
   ^4+
  39 \mid z \mid ^2 +6 \right)}
   {\left(1- \mid z \mid ^2\right)^2
   \left(9\mid z \mid ^4 +14 \mid z \mid ^2 +1\right)}$\\
   \hline
   &&\\
  8 &
  $-\mid z \mid^2\frac{\mid z \mid^6+49\mid z \mid^4+115\mid z \mid^2+27}
  {(1-\mid z \mid^4)(\mid z \mid^4+22\mid z
  \mid^2+1)}$&$\mid z \mid^2\frac{6(9\mid z \mid^4+62\mid z \mid^2+9)}
  {(1-\mid z \mid^2)^2(\mid z \mid^4+22\mid z
  \mid^2+1)}$\\
  \hline
\end{tabular}

\subsection{ Uncertainty Relations}

The uncertainty relations read

\begin{equation}\label{rel-incJ}
<(\Delta J_1)^2 >_z {}<(\Delta J_2)^2 >_z
\geq \frac{1}{4}  {} < J_0>_z^2,\quad  circ.perm..
\end{equation}
where the fluctuation of the operator T is defined by $\Delta T
= T -<T>_z$. From (\ref{Jx2-ini}), we get
:

\begin{equation}\label{rel-incxy}
<(\Delta J_1)^2 >_z {}<(\Delta J_2)^2 >_z
= \frac{1}{4} {}< J_z >_z ^2
\end{equation}
The M.F.S. minimize one of the uncertainty relations. The two
others are minimum if

\begin{equation}\label{rel-incxz}
  \begin{array}{ll}
&0=< (\Delta J_1)^2 >_z {}<(\Delta J_0)^2 >_z = < (\Delta J_2)^2
>_z  {}<(\Delta J_0)^2 >_z
=\\
&= \frac{(z\overline{z'})^2 N'(z\overline{z'})}{2 (N(z\overline{z'}))^3}
\left(N(z\overline{z'})(z\overline{z'}N''(z\overline{z'})+N'(z\overline{z'}))-z\overline{z'}
(N'(z\overline{z'}))^2\right)
  \end{array}
\end{equation}
that gives $x N(x) N''(x) -x (N'(x))^2 + N(x)N'(x)=0$. The solution
of this equation of the form (\ref{cond1-cj}) is the monomial
$x^l$, with $2 l \in N$. It results that the sequence $c_j$ is
restricted to one element $c_j
=
\delta_{jl}$ and that  the M.C.S.  span $h_l$,
 the resolution of unity does not exist in ${\cal{H}}_\frac{1}{2}$
  (or ${\cal{H}}_1$).

{\bf Result : }The M.F.S. minimize one, and only one, of the
uncertainty relations, this property is independent of the choice
of the sequence $c_j$.

\subsection{ Equation of motion}

When the evolution is defined by the hamiltonian (\ref{rotor}), the
molecular-components of the angular momentum, in the Heisenberg
representation, are time-dependent whereas their mean values on the
M.F.S. are
 time-independent.
The mean values of the angular momentum
  do not evolve as
 the  classical angular velocity of the rotor.

\subsection{ Expectation values of  the bi-spinor $S$ and of the bi-vector $V$}

Using the representation (\ref{reprS--}) and (\ref{reprS++}), we
easily prove that $< S_{-+}>_z$ and $ < S_{+-}>_z$ are equal to $0$
for all sequence $c_j$ and that

\begin{equation}\label{<S-->}
 <z\mid S_{--}\mid z> = \sum_{j} \overline{c}_{j+\frac{1}{2}} c_j \overline{z}^{j+\frac{1}{2}} z^j
     \sqrt{\frac{2j+1}{2j+2}}
= \overline{<z\mid S_{++}\mid z>}
\end{equation}

{\bf Result : } The $<S_{qq'}>_z$ form a $2\times 2$
diagonal-matrix that reduces to the $0$-matrix when the
representation space is ${\cal H}_1$, this result is independent on
the sequence $c_j$.

 Only the explicit expressions (\ref{<S-->}) of
the diagonal elements depend on the sequence. In the example 4, the
diagonal element are explicitly calculated
: $<S_{--}>_z = 2 \overline{z}^{\frac{1}{2}}\frac{1+2\mid
z\mid}{\mid z\mid^2+4\mid z\mid+1}$.

The mean values of the operators $V_{qq'}$ on the M.F.S. are
calculated by utilizing
 the expressions (\ref{reprV-+} $\cdots$ \ref{reprV00}), the
 non-diagonal elements $<V_{qq'}>_z$ are equal to $0$ and the
 diagonal elements are given by

\begin{equation}\label{<V00>}
  \begin{array}{ll}
    <z\mid V_{--}\mid z> &= \sum_{j} \overline{c}_{j+1} c_j \overline{z}^{j+1} z^j
     \sqrt{\frac{2j+1}{2j+3}}\\
     &\\
     & = \overline{<z\mid V_{++}\mid z>}\\
     &\\
<z\mid V_{00}\mid z> &= -\sum_{j} \frac{j}{j+1}\mid c_j \mid^2 \mid z\mid^{2j}\\
\end{array}
\end{equation}

{\bf Result : } $<V>_z$ is a $3\times 3$ diagonal-matrix, two of
the diagonal elements are complex conjugate and the third one is
real; this result does not depend on the sequence $c_j$.

 Only the
explicit expressions of the diagonal elements depend on the choice
of the sequence. In the examples 4 and 8, the calculations of
(\ref{<V00>}) give

\begin{equation}
\begin{tabular}{|l|l|l|} \hline
   & $<V_{--}>_z$ & $<V_{00}>_z$ \\
    \hline
     &&\\
  4 &
   $\overline{z} \frac{-\mid z\mid^2+4\mid z\mid+3}{\mid z\mid^2+4\mid z\mid+1}$
    & $\frac{\mid z\mid^2(-2\mid z\mid^3+\mid z\mid^2-6\mid z\mid+1^)
    -2\left(\log(1-\mid z\mid)-\mid z\mid\right)}{\mid z\mid^2(1-\mid z\mid)^4}$\\
     &&\\
     \hline
      \hline
       &&\\
      8 & $\overline{z} \frac{-\mid z\mid^8+8\mid z\mid^6+
  110\mid z\mid^4+240\mid z\mid^2+27}{(1-\mid z\mid^4)
  (\mid z\mid^4+22\mid z\mid^2+1)}$ &$ \frac{-19\mid z\mid^8+5\mid z\mid^6
  -41\mid z\mid^4+7\mid z\mid^2
  -(1-\mid z\mid^2)^4 \log(1-\mid z\mid^2)}{\mid z\mid^2(1+\mid z\mid^2)
  (\mid z\mid^4+22\mid z\mid^2+1)}$\\
  &&\\
 \hline
\end{tabular}
\end{equation}

Similar results hold for bi-tensors of higher equal rank.

To end, let us stress that it is difficult to study the uncertainty
relations between the angular momentum and the bi-tensors $S$ and
$V$, and in particular to find the sequence $c_j$ that minimizes
any of them.

In the following section, we show that all the properties of the
M.F.S. have a counterpart for the M.C.S.

\section{Properties of the Coherent States}\label{prop-CS}

\subsection{ Action of the angular momentum on the M.C.S.}

First, we remark that a M.C.S. is not transformed into a M.C.S. by
the components of $\vec{J}$. The operator
$D_L(\zeta_L)D_M(\zeta_M)$ transforms the angular momentum
$\vec{J}$ in a vector, the laboratory and molecular-components of
which are ~:
\begin{equation}\label{Jtransform}
D_L(\zeta_L)J^L_{q'} D^{-1}_L(\zeta_L)\equiv J^{L}_{q'}(\zeta_L),
\quad D_M(\zeta, M)J^M_{q} D^{-1}_M(\zeta_M)\equiv J^{M}_{q}(\zeta_M)
\end{equation}
From (\ref{eta}) and (\ref{prop-ini1}), we obtain~:

\begin{equation}\label{prop1}
J^{L}_{0}(\zeta_L)\mid Z>
 = J^{M}_{0}(\zeta_M)\mid Z>
 = -\Lambda \mid Z>
\end{equation}
Let us remark that the action of the operators $J^{L}_{0}(\zeta_L)
$ and of $J^{M}_{0}(\zeta_M)$ transforms the set of M.C.S.
associated to the sequence $c_j $ into the set of M.C.S. associated
to the sequence $j c_j$ and that these two sets of M.C.S.
correspond to the same domain of the complex plane $z$. Formula
(\ref{prop1}) can be written on the form~:

\begin{equation}\label{nJL}
  \begin{array}{ll}
    (\frac{1}{2}e^{-i\varphi_L}\sin\theta_L J_+^L+
    \cos\theta_L J_0^L+ \frac{1}{2}e^{i\varphi_L}\sin\theta_L
    J_-^L)\mid Z>
    &= -\Lambda \mid Z> \\
    =(\cos\varphi_L \sin\theta_L J_1^L
    +\sin\varphi_L \sin\theta_L J_2^L+\cos \theta_L J_0^L)\mid Z>& =
    (\vec{n}^{ L}(\zeta_L).\vec{J})\mid Z>
  \end{array}
\end{equation}
An analogous relation holds for the molecular-components of
$\vec{J}$. We write these relations in a more compact form~:

\begin{equation}\label{nJ}
 \left(\Lambda + \vec{n}^{ M}(\zeta_M).\vec{J}\right)\mid Z> =
 \left(\Lambda +\vec{n}^{ L}(\zeta_L).\vec{J}\right)\mid Z> = 0
\end{equation}

The operator $(\vec{J}.\vec{n}^{ L}(\zeta_L))$ (resp. $\vec{n}^{
M}(\zeta_M).\vec{J}$) is the projection of the angular momentum
$\vec{J}$ on the vector $\vec{n}^{ L}(\zeta_L)$ (resp. $\vec{n}^{
M}(\zeta_M)$) , the laboratory-components(resp.
molecular-components) of which are $(\cos\varphi_L
\sin\theta_L$, $
\sin\varphi_L \sin\theta_L$, $\cos \theta_L)$ ( resp.
$(\cos\varphi_M
\sin\theta_M$, $
-\sin\varphi_M \sin\theta_M$, $\cos \theta_M)$). This vector
 $\vec{n}^{ L}(\zeta_L)$ (resp. $\vec{n}^{ M}(\zeta_M)$)
is the transformed of the unit vector  of the $x_0$-axis of the laboratory (
  resp. molecular) frame by $D_L(\zeta_L)$ (resp. $D_M(\zeta_M)$).

{\bf Result : } The projections of the angular momentum $\vec{J}$
on the two vectors $\vec{n}^{ L}(\zeta_L)$ and $\vec{n}^{
M}(\zeta_M)$ transform a M.C.S. belonging to some c-set into the
same M.C.S. that do not belong to this c-set.

From (\ref{prop-ini2}) we get

\begin{equation}\label{prop2}
  \begin{array}{lll}
J^{ L}_-(\zeta_L) \mid Z>
 &= (\zeta_L^2 J_+^L -2\zeta_L  J_0^L - J_-^L)\mid Z>
 &= 0\\
 &&\\
 J^{ M}_+ (\zeta_M) \mid Z>
     &= (\zeta_M^2 J_-^M -2\zeta_M  J_0^M - J_+^M)\mid Z> &= 0
      \end{array}
\end{equation}
The first relation was already obtained in \cite{perelomov} for the
spin C.S..

{\bf Remark :} All the relations obtained in this subsection are
independent on the sequence $c_j$.

\subsection{Laboratory and Molecular-rotations }

Let us put the product of the two laboratory-rotations on the form

\begin{equation}\label{RLDL1}
R_L(\alpha_L,\beta_L, \gamma_L) D(\zeta_L) =
D_L(R_L.\zeta_L)e^{i\lambda J^L_0}
\end{equation}

where
\begin{equation}\label{RLDL2}
 R_L.\zeta_L =
\frac{u_L \zeta_L + v_L}{\overline{
u_L}-\overline{v_L}\zeta_L} \mbox{and}\quad e^{i\lambda}= \left(
\frac {u_L
-v_L\overline{\zeta_L}}{\overline{
u_L}-\overline{v_L}\zeta_L}\right)
\end{equation}
We have denoted $u_L =
e^{-i\frac{\alpha_L+\gamma_L}{2}}\cos\frac{\beta_L}{2}$ and $ v_L =
e^{i\frac{\alpha_L-\gamma_L}{2}}\sin\frac{\beta_L}{2}$.

The action of the laboratory-rotation on the M.C.S.(\ref{eta})
result from (\ref{RLDL1})

\begin{equation}\label{TjZ}
\begin{array}{ll}
     R_L (\alpha_L,\beta_L, \gamma_L)\mid Z> & =
     \sum_j c_j z^j\left(
\frac {\overline{
u_L}-\overline{v_L}\zeta_L}{u_L
-v_L\overline{\zeta_L}}\right)^j D^j_L(R_L.\zeta_L)
 D^j_M(\zeta_M)\mid j, -j, -j>\\
 &\\
 &=  D_L(R_L.\zeta_L)D^j_M(\zeta_M)\mid R_L.z>
  \end{array}
\end{equation}
where

\begin{equation}\label{RLDL3}
  R_L.z = z\frac {\overline{
u_L}-\overline{v_L}\zeta_L}{u_L
-v_L\overline{\zeta_L}}
\end{equation}
 This result reads

\begin{equation}\label{RLZ}
   R_L (\alpha_L,\beta_L, \gamma_L)\mid z, \zeta_L, \zeta_M > =
   \mid R_L.z, R_L.\zeta_L, \zeta_M >
\end{equation}

Let us remark that

- the M.F.S. $\mid z> $ and $\mid R_L.z>$ correspond to the
 sequence $c_j$, the
laboratory-rotations $R_L (\alpha_L,\beta_L,
\gamma_L)$ transform a M.C.S. of a c-set
into a M.C.S. of the same c-set.

- due to the equality  $\mid z
\mid
= \mid R_L.z \mid$,
 the norms of the two M.C.S. $\mid Z>$ and $R_L\mid Z>$ are equal.

Obviously, we get the analogous result for the action of
molecular-rotations $R_M$ that act in one c-set according to the
formula :

\begin{equation}\label{RMZ}
   R_M (\alpha_M,\beta_M, \gamma_M)\mid z, \zeta_L, \zeta_M > =
   \mid R_M.z, \zeta_L,R_M.\zeta_M >
\end{equation}
where the transformed complex variables $R_M.z$ and $R_M.\zeta_M$
are given by the formulas obtained by replacing the label
 $L$ by  $M$ in (\ref{RLDL2}) and (\ref{RLDL3}). The
 molecular-rotations play a crucial part in the
 study of the symmetry of the molecule that will be the subject of a
  forthcoming paper.

\subsection{Expectation values of the angular momentum}

The rotations transform a bi-tensor $T^{jj'}_{qq'}$ into the
operators

\begin{equation}\label{DT}
D_L(\zeta_L)D_M(\zeta_M)
    T^{jj'}_{qq'}D^{-1}_L(\zeta_L)D_M^{-1}(\zeta_M)
    \equiv T^{ jj'}_{qq'}(\zeta_M,\zeta_L)
\end{equation}
that satisfy the commutation relations (\ref{J'T}) in which the
angular momentum is replaced by the transformed angular momentum
(\ref{Jtransform}). From(\ref{tens-R}) and (\ref{eta}), one deduces

\begin{equation}\label{<ztensz1>}
\begin{array}{ll}
    <z\mid T^{jj'}_{qq'}\mid z> & = <Z \mid T^{ jj'}_{qq'}(\zeta_M,\zeta_L)\mid Z>\\
     & = \sum_{k'=-j'}^{j'}\sum_{k=-j}^{j}\tilde{D}^j_{qk}(\zeta_M)
     <Z\mid T^{jj'}_{kk'}\mid Z> D^{j'}_{k'q'}(\zeta_L)\\
\end{array}
\end{equation}
Let $< T >_Z$ denote the expectation value of $T$ in the state
$\mid Z>$, namely $\frac{ <Z\mid T\mid Z>}{ <Z\mid Z>}$. We have

\begin{equation}\label{<ztensz2>}
< T >_z = <T(\zeta_M, \zeta_L)>_Z = \tilde{D}(\zeta_M)< T >_Z
D(\zeta_L)
\end{equation}

\vspace{0.5cm}

Let us apply the previous result to the case of the bi-tensor
$T^{01}$ and $T^{10}$. From (\ref{<ztensz2>}) and (\ref{<J>}), we
deduce that
:

$\clubsuit$ The vector
  $< J^{ L}_{0,1,2}(\zeta_L)>$ lies along the
  $x_0$-axis of the laboratory-frame. Analogously,
 the vector $< J^{M}_{0,1,2}(\zeta_M)>$ lies along the
  $x_0$-axis of the molecular-frame.

$\clubsuit$ Remembering that $(-\frac{1}{\sqrt{2}}J^L_+, J^L_0,
\frac{1}{\sqrt{2}}J^L_-) $ is a bi-tensor $J^{0 1}_{0q'}$,
the calculation of $<Z\mid J^L_{q'}\mid Z>$ is done by using
(\ref{<ztensz2>}), we obtain

\begin{equation}\label{JLZ}
\begin{array}{ll}
    <Z \mid J^{0 1}_{q'}\mid Z> & = \sum^1_{-1} <z \mid J^{0 1}_{k'}\mid z>
    D^1_{k' q'}(-\zeta_L) \\
    &\\
   < J^{0 1}_{q'}>_Z  & =  <  J_{0}>_z D^1_{0 q'}(-\zeta_L)
  \end{array}
\end{equation}

We verify that the vector $< J_{q'}^L>_Z$ is parallel to the
previously introduced vector $n^L_{q'}(\zeta_L)= D^1_{0
q'}(-\zeta_L)$. This result was obtained in \cite{janssen} and
\cite{mdo}, but it is interesting to point out that this result
holds for any c-set of M.C.S. Only the length of the vector depends
on the choice of coefficients $c_j$ and of the value of $z$.

$\clubsuit$ Similarly, the calculation of $<Z\mid J^M_{q'}\mid Z>$
is performed by applying (\ref{<ztensz2>}) to the bi-tensor $J^{1
0}_{q0} =(-\frac{1}{\sqrt{2}}J^M_-, J^M_0,
\frac{1}{\sqrt{2}}J^M_+)$

\begin{equation}\label{JMZ}
\begin{array}{ll}
    <Z\mid J^{ 10}_{q}\mid Z> & = \sum^1_{-1}
    <z\mid J^{ 1 0}_{k}\mid z> D^1_{k q}(-\zeta_M) \\
    &\\
   < J^{ 10}_{q}>_Z  & =  < J_{0}>_z D^1_{0 q}(-\zeta_M)
  \end{array}
\end{equation}
The vectors $< J^M>_Z$ and $\vec{n}^M (\zeta_M)$ are parallel. This
result does not depend on the c-set considered.

Let us remark that the norms of the vectors $< \vec{J}^M>_Z$ and
 $< \vec{J}^L>_Z$ are both equal to the absolute value of
$< J_{0}>_z = -\mid z\mid^2 \frac{N'(\mid z\mid^2)}{N(\mid
z\mid^2)}$.
The expectation values of the Casimir operator $J^2$ on the M.F.S.
and on the M.C.S. are equal, from (\ref{J}), we have~:

\begin{equation}\label{J2Z}
  <J^2 >_Z =   \mid z \mid^2 \left( \mid z \mid^2
     \frac{N''(\mid z \mid^2)}{N(\mid z \mid^2)}+
     2 \frac{N'(\mid z \mid^2)}{N(\mid z \mid^2)}\right)
\end{equation}

{\bf Result :} Interpretation of $Z$

The angles $\theta_L$ and $\varphi_L$ (resp. $\theta_M$ and
$\varphi_M$) define the direction of the vector $< J^L_{q'}>_Z$
(resp. $< J^M_q>_Z$) in the laboratory(resp. molecular) frame. The
modulus $\mid z \mid$ and the choice of the $c_j$ are related to
the length of these vectors and to the mean values of $J^2$.
\subsection{ Uncertainty Relations}

The transformed operators $ J^{ L}_i(\zeta_L)$ and $ J^{
M}_i(\zeta_M)$ satisfy the same commutation relations (\ref{com'})
and (\ref{com}) as $J^L_i$ and $J^M_i$. Therefore their
fluctuations obey the same inequalities (\ref{rel-incJ}). We easily
verify that $<(\Delta
\vec{J})^2>_z
= <(\Delta
\vec{J}(\zeta))^2>_Z$. From the equality
(\ref{rel-incxy}), we establish that the M.C.S. minimize two
uncertainty relations, namely

\begin{equation}\label{ZL-rel-incxy}
<(\Delta J^{ L}_1(\zeta_L))^2>_Z <(\Delta J^{ L}_2(\zeta_L))^2>_Z
= \frac{1}{4} \left(< J^{ L}_0(\zeta_L)>_Z\right)^2
\end{equation}
and

\begin{equation}\label{ZM-rel-incxy}
<(\Delta J^{ M}_1(\zeta_M))^2>_Z <(\Delta J^{ M}_2(\zeta_M))^2>_Z
= \frac{1}{4} \left(< J^{ M}_0(\zeta_M)>_Z\right)^2
\end{equation}
The M.C.S. do not minimize the uncertainty relations involving the
$J^L_i$ and $J^M_i$, as studied by \cite{janssen} \cite{mdo}, but
verify two equalities (\ref{ZL-rel-incxy}) and (\ref{ZM-rel-incxy})
involving the transformed operators $ J^{ L}_i(\zeta_L)$ and $ J^{
M}_i(\zeta_M)$. A similar result occurs for the spin
C.S.\cite{perelomov}.

\subsection{ Expectation values of the bi-spinor and the
 bi-vector}

It results from Formula (\ref{<ztensz2>}) that the two matrices
$<S(\zeta_M,\zeta_L)>_Z$ and $<V(\zeta_M,\zeta_L)>_Z$ are diagonal
and that

\begin{equation}\label{ZSZ1}
  <Z\mid S_{q q'}\mid Z>  = \sum_{k'=(-\frac{1}{2},\frac{1}{2})}
    \sum_{k
    =(-\frac{1}{2},\frac{1}{2})}\tilde{R}^\frac{1}{2}_{qk}(-\zeta_M)
      <z\mid S_{k k'}\mid z> R^\frac{1}{2}_{k'q'}(-\zeta_L)
\end{equation}
The matrix $<S>_Z$ then is the product of one matrix
$\tilde{R}^\frac{1}{2}(-\zeta_M)$ only depending on $\zeta_M$, one
diagonal matrix $<S>_z$ only depending on the coefficients $c_j$
and on $z$, and one matrix $R^\frac{1}{2}(-\zeta_L)$ only depending
on $\zeta_L$.

Similarly, applying the formula(\ref{<ztensz2>}) to the bi-vector,
we get

\begin{equation}\label{ZVZ1}
  <Z\mid V_{q q'}\mid Z>  = \sum_{k'=(-1,1)}
    \sum_{k
    =(-1, 1)}\tilde{R}^1_{qk}(-\zeta_M)
      <z\mid V_{k k'}\mid z> R^1_{k'q'}(-\zeta_L)
\end{equation}
Therefore the matrix $<V>_Z$ is the product of
$\tilde{R}^1(-\zeta_M)$ by the diagonal matrix $<V>_z$ and by
$R^1(-\zeta_L)$, $<V>_z$ depends on the coefficients $c_j$ and on
$z$.

These results can be extended to bi-tensors of higher equal rank.

In conclusion, we have obtained the decomposition of the bi-spinor
matrix $<S>_Z$ and of the bi-vector matrix $<V>_Z$ in terms of
three matrices, each of these matrices only depends on one complex
variable $z$, $\zeta_L$ or $\zeta_M$.

\subsection{ Evolution equation}

$\clubsuit$ {\bf Rotor} Let us consider a quantum rigid molecule
described by the hamiltonian (\ref{rotor}). In the Schr\"{o}dinger
representation, the evolution of the M.C.S. $\mid Z>$ is given by
$e^{iHt}\mid Z>$ that obviously is not a M.C.S. A top in a M.C.S.
does not remain in a M.C.S. All the demonstrations of
\cite{janssen} and \cite{mdo}, based on the properties of the
M.C.S. are not valid at a time $t
\ne 0$. In particular, the expectation values  $<J_iJ_k +J_kJ_i>$ are not equal
 to $2<J_i><J_k>$ when $t \ne 0$ and the evolution equations of the expectation
values of the angular momentum are not classical.

Remarks : For the spherical rotor ($A_1
= A_2
= A_3$),

- the  M.C.S. corresponding to a sequence $c_j$ become the M.C.S.
corresponding to a sequence $c_j e^{it\frac{j(j+1)}{A}}$
 during the motion,

- the expectation
values of all the components of the angular momentum are constant
and then correspond to the classical rotation vector.

 - Following \cite{kl},  we obtain  M.C.S.
  that have the temporal stability by replacing
  $z^j$ by $\mid z\mid^j e^{j\tau(j+1)}$ in (\ref{initial}).

\noindent $\clubsuit$ {\bf Temporal stability of (\ref{eta})}

We look for an Hamiltonian $H_{\natural}$ such that, in the
Schr\"odinger representation, the system is described by the state
$\mid Z(t)>
\equiv \mid z(t),
\zeta_L(t), \zeta_M(t)>$. Using the definition (\ref{eta}) of the M.C.S.
and the expressions of the components of $\vec{J}$ given in Section
(\ref{Z-Representation}), we prove that the evolution equation of
the state is of the form :

\begin{equation}\label{eq-2}
\begin{array}{ll}
& i\partial_t \mid Z(t)> = H_{\natural}\mid Z(t)>\\ &=
\left(i(a^LJ^L_+-\overline{a}^LJ^L_-) +
a^L_0J^L_0+i(a^MJ^M_+-\overline{a}^MJ^M_-) + a^M_0J^M_0\right)\mid
Z(t)>
\end{array}
\end{equation}
The complex variables $\zeta_L(t)$ and $\zeta_M(t)$ defining the
M.C.S. are related to the coefficients $a$ occurring in the
hamiltonian $H_{\natural}$~:

\begin{equation}\label{zetaL-t}
\dot{\zeta}_L(t) = a^L + \overline{a}^L\zeta^2_L(t)-i a^L_0
\zeta_L(t),
\end{equation}
and

\begin{equation}\label{zetaM-t}
\dot{\zeta}_M(t) = a^M + \overline{a}^M\zeta^2_M(t)-i a^M_0 \zeta_M(t)
\end{equation}
The complex variable $z(t)$ must be of the form $z e^{-i\sigma(t)}$
in order that the $H_{\natural}$ be hermitean and $\sigma(t)$ must
verify

\begin{equation}\label{sigma-t}
\dot{\sigma}(t) = i\left(a^L \overline{\zeta}_L(t)-
\overline{a}^L\zeta_L(t)\right) - a^L_0 +i\left(a^M \overline{\zeta}_M(t)-
\overline{a}^M\zeta_M(t)\right) - a^M_0
\end{equation}
Writing $H_{\natural}$ on the form $\sum
_i h^L_iJ^L_i$, we deduce that this equation describes the
motion of a rigid body in a magnetic field $h^L_i$, that depends on
the time through the coefficients $a$ and of the position of the
rigid body in the laboratory through the molecular-components
$J^M$.

Let us stress that the state $ \mid Z(t)>$ being a M.C.S., the
expectation values of the components of $\vec{J}$ take the form
(\ref{JLZ}) and (\ref{JMZ}):

\begin{equation}\label{evol1}
< Z( t) \mid J^L_i \mid Z( t)> = < J_{0}>_z \nu^L_i(t), \quad < Z(
t)
\mid J^M_i \mid Z( t)> = < J_{0}>_z \nu^M_i(t)
\end{equation}
where the vectors $\vec{_nu}^L(t) $ and $\vec{\nu}^M(t)$ verify
classical equations of motion. This generalizes the result of
Perelomov \cite{perelomov1} for $su(2)$.

In the previous reasoning, the $c_j$ are time-independent. When the
$c_j$ depend on $t$, the hamiltonian $H_{\natural}$ contains an
extra term that is a function of the Casimir operator $J^2$ and of
$t$.

\section{Conclusion}

  Molecular-Coherent-States are constructed by transforming
  Molecular-Fundamental-States by
   laboratory and  molecular-rotations. A M.F.S. is assumed to be
   a
   linear
    combination of the form (\ref{initial}) in
     which the coefficients $c_j$ have to
    verify two conditions in order that the M.C.S. satify the
    properties~:

$\bullet $ 1) The M.C.S. constitute an (overcomplete) basis of non
orthogonal vectors of ${\cal H}_\frac{1}{2}$ (or eventually ${\cal
H}_1$).

$\bullet $ 2) The vectors of ${\cal H}_\frac{1}{2}$ are realized as
continuous functions of three complex variables in section
(\ref{Z-Representation}).

We have established the list of properties of the M.C.S., in
analogy
 to that given in the introduction for the C.S.H.O~:

$\bullet $ 3) The four operators defined in (\ref{nJ}) and
(\ref{prop2}) transform the M.C.S. into $0$.

$\bullet $ 4) The M.C.S. minimize two uncertainty relations
(\ref{ZL-rel-incxy}) and (\ref{ZM-rel-incxy}).

$\bullet $ 5) The expectation values of the components of the
angular momentum evolve classically for a molecule in a magnetic
field.

$\bullet $ 6) For such a quantum system, the temporal stability is
verified. When time evolves, a M.C.S. remains a M.C.S.. However,
this is not true for the top-hamiltonian (\ref{rotor}) contrary to
what was claimed by \cite{janssen} and \cite{mdo}.

$\bullet $ 7) By construction, the M.C.S. are generated by the
group of laboratory and molecular-rotations.

We have seen that the prominent part in the group-construction of
the M.C.S. is played by the M.F.S.. All the calculations involving
M.C.S. are reduced to simpler ones involving M.F.S.. In particular,
 we easily establish that the matrices
of the expectation values of the bi-spinor and the bi-vector are
decomposed into the product of two rotations and a diagonal matrix.

       To conclude, let us stress the following results~:

-the choice of the M.F.S. is the only arbitrariness of the
group-construction of the M.C.S.,

- the fact that the M.F.S.
       are expressed in terms of the vectors $\mid j,-j,-j>$ play a crucial part
        in the establishment of all the
properties,

 - these properties are true for any sequence
$c_j$, and we were not able to distinguish and choose a specific
sequence and then a more prominent M.F.S.. Therefore, we can choose
in each problem the more convenient basis.

\section{Appendix}
\subsection{Realization in ${\cal{C}}(\alpha,
\beta,\gamma)$}\label{ap-Euler}

In the space of the functions of the Euler angles, ${\cal{C}}(\alpha,
\beta,\gamma)$,
 the
laboratory and the molecular-components of the angular momentum
take the form~:

\begin{equation}
  \begin{array}{ll}
    J^L_{+}& = i \exp(i\alpha)\left(  \cot \beta \partial_\alpha
     -i \partial_\beta -
    \frac{1}{\sin \beta}\partial_\gamma\right)\\
     J^L_{-}& = i \exp(-i\alpha)\left(  \cot \beta \partial_\alpha
     +i \partial_\beta -
    \frac{1}{\sin \beta}\partial_\gamma\right)\\
    J^L_{0}& =-i\partial_\alpha
  \end{array}
\end{equation}

and

\begin{equation}
  \begin{array}{ll}
    J^M_{+}& = -i \exp(-i\gamma)\left(  \cot \beta \partial_\gamma
     +i \partial_\beta -
    \frac{1}{\sin \beta}\partial_\alpha\right)\\
     J^M_{-}& = -i \exp(i\gamma)\left(  \cot \beta \partial_\gamma
     -i \partial_\beta -
    \frac{1}{\sin \beta}\partial_\alpha\right)\\
    J^M_{0}& =-i\partial_\gamma
  \end{array}
\end{equation}
The states $\mid j, k, m >$ are represented by the functions

\begin{equation}
<\alpha,\beta,\gamma \mid j k m > =
R^{*j}_{mk}(\alpha,\beta,\gamma)=
\sqrt{2j+1} \exp(im\alpha)\exp(ik\gamma)d^{j}_{mk}(\beta)
\end{equation}
where the $R^j_{mk}$ were defined in (\ref{Djmm'1}) and
(\ref{Djmm'2}). The functions are singled valued if $j, k$ and $ m$
are integer numbers.

A M.F.S. is realized as a function of the variable $z
e^{-i(\alpha+\gamma)}\cos^2\frac{\beta}{2}$.

\subsection{Representation of $S$ and $V$}\label{S-V}

Using the formulas (\ref{J'T}) and (\ref{reprJJ'}), we obtain the
actions of all the components of $S$ on the canonical basis from
one of them. We get~:

\begin{equation}\label{reprS++}
\begin{array}{ll}
  S_{+ \pm }\mid j, k, m >
  &= \frac{\sqrt{(j \pm m+1)(j+k+1)}}{\sqrt{2(j+1)(2j+1)}}
  \mid j+\frac{1}{2}, k+\frac{1}{2}, m \pm \frac{1}{2} >\\
&\pm \frac{\sqrt{(j \mp m)(j-k)}}{\sqrt{2j(2j+1)}}
  \mid j-\frac{1}{2}, k+\frac{1}{2}, m \pm \frac{1}{2} >,\\
\end{array}
\end{equation}

\begin{equation}\label{reprS--}
\begin{array}{ll}
  S_{- \pm }\mid j, k, m >
  &= \frac{\sqrt{(j \pm m+1)(j-k+1)}}{\sqrt{2(j+1)(2j+1)}}
  \mid j+\frac{1}{2}, k-\frac{1}{2}, m \pm \frac{1}{2} >\\
&\mp \frac{\sqrt{(j \mp m)(j+k)}}{\sqrt{2j(2j+1)}}
  \mid j-\frac{1}{2}, k-\frac{1}{2}, m \pm \frac{1}{2} >,\\
\end{array}
\end{equation}

The components of $V$ act on the canonical basis according to the
formulas~:

\begin{equation}\label{reprV-+}
\begin{array}{ll}
  V_{-\pm}\mid j, k, m >
  &= \frac{\sqrt{(j\pm m+1)(j\pm m+2)(j-k+1)(j-k+2)}}{2(j+1)\sqrt{(2j+1)(2j+3)}}
  \mid j+1, k-1, m\pm 1 >\\
&\mp \frac{\sqrt{(j\mp m)(j\pm m+1)(j-k+1)(j+k)}}{2(j+1)j} \mid j,
k-1, m\pm 1>\\ &+\frac{\sqrt{(j\mp m-1)(j\mp
m)(j+k-1)(j+k)}}{2j\sqrt{(2j+1)(2j-1)}}\mid j-1, k-1, m\pm 1>
\end{array}
\end{equation}

\begin{equation}\label{reprV++}
\begin{array}{ll}
 V_{+\pm}\mid j, k, m >
  &= \frac{\sqrt{(j\pm m+1)(j\pm m+2)(j+k+1)(j+k+2)}}{2(j+1)\sqrt{(2j+1)(2j+3)}}
  \mid j+1, k+1, m\pm 1 >\\
&\pm \frac{\sqrt{(j\pm m+1)(j\mp m)(j+k+1)(j-k)} }{2(j+1)j} \mid j,
k+1, m\pm 1>\\ &+\frac{\sqrt{(j\mp m-1)(j\mp
m)(j-k-1)(j-k)}}{2j\sqrt{(2j+1)(2j-1)}}\mid j-1, k+1, m\pm 1>
\end{array}
\end{equation}

\begin{equation}\label{reprV+0}
\begin{array}{ll}
  V_{\pm 0}\mid j, k, m >
  &= \frac{ \sqrt{(j-m+1)(j+m+1)(j\pm k+1)(j\pm k+2)}}
  {\sqrt{2}(j+1)\sqrt{(2j+1)(2j+3)}}
  \mid j+1, k\pm 1, m>\\
&\mp\frac{m \sqrt{(j\mp k)(j\pm k+1)}}{\sqrt{2}(j+1)j}\mid j, k\pm
1, m>\\ &-\frac{\sqrt{(j-m)(j+m)(j\mp k-1)(j\mp
k)}}{\sqrt{2}j\sqrt{(2j+1)(2j-1)}} \mid j-1, k\pm 1, m>
\end{array}
\end{equation}

\begin{equation}\label{reprV0+}
\begin{array}{ll}
  V_{0\pm}\mid j, k, m >
  &= \frac{\sqrt{(j\pm m+1)(j\pm m+2)(j-k+1)(j+k+1)}}
  {\sqrt{2}(j+1)\sqrt{(2j+1)(2j+3)}}
  \mid j+1, k, m\pm 1 >\\
&\mp\frac{k \sqrt{(j\mp m)(j\pm m+1)}}{\sqrt{2}(j+1)j} \mid j, k,
m\pm 1>\\ &-\frac{\sqrt{(j\mp m-1)(j\mp
m)(j-k)(j+k)}}{\sqrt{2}j\sqrt{(2j+1)(2j-1)}}\mid j-1, k, m\pm 1>
\end{array}
\end{equation}

\begin{equation}\label{reprV00}
\begin{array}{ll}
 V_{00}\mid j, k, m >
  &= \frac{\sqrt{(j-m+1)(j+m+1)(j-k+1)(j+k+1)}}{(j+1)\sqrt{(2j+1)(2j+3)}}
  \mid j+1, k, m>\\
&+ \frac{m k}{(j+1)j} \mid j, k, m>\\
&+\frac{\sqrt{(j-m)(j+m)(j-k)(j+k)}}{j\sqrt{(2j+1)(2j-1)}}\mid j-1,
k, m>
\end{array}
\end{equation}



\begin{thebibliography}{99}
\bibitem{schrodinger} E.
Schr\"odinger,
"Der Stetige Ubergang von der Mikro-zur
Makromechanik" Naturwissenschaften, 14, (1926), 664.
 \bibitem{Klauder} John R. Klauder and Bo-Sture Skagerstam,
 Coherent States : Applications in Physics and
Mathematical Physics" (World Scientific, 1985).
 \bibitem{perelomov}
 A.M.Perelomov, "Generalized Coherent States and Their Applications"
 (Springer-Verlag, 1986).
 \bibitem{perelomov1}  A.M.Perelomov,
 "Coherent States for Arbitrary Lie Group",
 Commun.math.Phys. 26, (1972), 222.
 \bibitem{radcliffe} J.M.
Radcliffe,
"Some properties of coherent spin states" J.Phys.A :
Gen.Phys., 4, (1971) 270.
\bibitem{arecchi} F.T.Arecchi, Eric Courtens, Robert Gilmore and Harry Thomas,
 "Atomic Coherent States in Quantum Optics",
Physical Review A,Vol. 6,No. 6, (1972) 2211.
\bibitem{janssen}  D.Janssen,
"Coherent states of the quantum-mechanical top",
Sov. J. Nucl.Phys. 25, 4, (1977),479.
\bibitem{mdo}  Jorge A.
 Morales, Erik Deumens and Yngve \"{O}hrn,
 "On rotational coherent states in molecular quantum dynamics"
 J. Math. Phys. Vol.
 40, No. 2, (1999), 766.
\bibitem{kl} John R. Klauder,
"Coherent states without groups: quantization on nonhomogeneous
manifold" Modern Physics Letters A, 8, 18 (1993) 1735.
\bibitem{atkins} P.W.Atkins and
J.C.Dobson,
"Angular momentum coherent states" Proc.Roy.Soc.Lond.A.
321 (1971) 321.
\bibitem{bhaumik}  Debajyoti
Bhaumik, Tarashankar Nag and Binayak Dutta-Roy,
"Coherent states for
angular momentum" J.Phys.A : Math.Gen., 8, 12 (1975) 1868.
\bibitem{fonda} Luciano Fonda, Norma Manko\u{c}- Bor\u{s}tnik and Mitja Rosina,
"Coherent Rotational States, Their Formation and Detection" Physics
Reports 158, 3 (1988) 159.
\bibitem{pavlichenkov} I.M.Pavlichenkov,
 "Quantum theory of the asymmetric top"
Sov.J.Nucl.Phys. 33, 1, (1981) 52.
\bibitem{takahashi} Yoshinori Takahashi and
 Fumiaki Shibata,
 "Spin Coherent State Representation in
 Non-Equilibrium Statistical Mechanics", Journal of The Physical Society of Japan, Vol.
 38, 3, (1975), 656.
\bibitem{judd} B.R.Judd,
"Angular Momentum Theory for Diatomic Molecules"
(Academic Press, New-York, 1975).
\end{thebibliography}
\end{document}